%% file: iros_2022_sekiguchi.tex
\documentclass[letterpaper, 10 pt, conference]{ieeeconf}
\usepackage[utf8]{inputenc}

\IEEEoverridecommandlockouts

\usepackage[pdftex]{graphicx}

\usepackage{amssymb,amsmath,amsfonts}
\usepackage{booktabs}
\usepackage{multirow}
\usepackage{bm}
\usepackage{cite}
\usepackage{mathtools}
\usepackage{microtype}
\usepackage{url}
\usepackage{xcolor}
\usepackage{comment}

\usepackage{color}

\input{definition}

\title{\LARGE \bf
Direction-Aware Adaptive Online Neural Speech Enhancement 
with an Augmented Reality Headset 
in Real Noisy Conversational Environments
}

\author{
    Kouhei Sekiguchi$^{1}$,
    Aditya Arie Nugraha$^{1}$,
    Yicheng Du$^{2}$, \\
    Yoshiaki Bando$^{3}$,
    Mathieu Fontaine$^{4}$,
    and Kazuyoshi Yoshii$^{5}$%
    \thanks{
        This work was supported by 
        JSPS KAKENHI Nos.~19H04137, 20K19833, 20H01159, and 20K21813.}%
    \thanks{
        $^{1}$\,
        Kouhei Sekiguchi and Aditya Arie Nugraha are with
        the Center for Advanced Intelligence Project (AIP), RIKEN, Tokyo, Japan
        (e-mail: \protect\url{adityaarie.nugraha@riken.jp}, \protect\url{kouhei.sekiguchi@riken.jp}).
        These authors are co-first authors contributing equally to this work.}%
    \thanks{
        $^{2}$\,
        Yicheng Du is with 
        the Graduate School of Informatics, Kyoto University, Kyoto, Japan
        (e-mail: \protect\url{du.yicheng.64z@st.kyoto-u.ac.jp}).}%
    \thanks{
        $^{3}$\,
        Yoshiaki Bando is with
        the National Institute of Advanced Industrial Science and Technology (AIST), Tokyo, Japan and also with
        the Center for Advanced Intelligence Project (AIP), RIKEN, Tokyo, Japan
        (e-mail: \protect\url{y.bando@aist.go.jp}).}%
    \thanks{
        $^{4}$\,
        Mathieu Fontaine is with 
        LTCI, Télécom Paris, Institut Polytechnique de Paris, Paris, France
        (e-mail: \protect\url{mathieu.fontaine@telecom-paris.fr}).}%
    \thanks{
        $^{5}$\,
        Kazuyoshi Yoshii is with 
        the Graduate School of Informatics, Kyoto University, Kyoto, Japan and also with
        the Center for Advanced Intelligence Project (AIP), RIKEN, Tokyo, Japan
        (e-mail: \protect\url{yoshii@i.kyoto-u.ac.jp}).}%
}

\begin{document}

\maketitle
\thispagestyle{empty}
\pagestyle{empty}

\begin{abstract}
This paper describes the practical response- and performance-aware development of online speech enhancement
 for an augmented reality (AR) headset
 that helps a user understand conversations made in real noisy echoic environments (\eg, cocktail party).
One may use a state-of-the-art blind source separation method 
 called fast multichannel nonnegative matrix factorization (FastMNMF) 
 that works well in various environments thanks to its unsupervised nature.
Its heavy computational cost, however, prevents its application to real-time processing.
In contrast, a supervised beamforming method
 that uses a deep neural network (DNN) for estimating spatial information of speech and noise
 readily fits real-time processing, but
 suffers from drastic performance degradation in mismatched conditions.
Given such complementary characteristics, 
 we propose a dual-process robust online speech enhancement method
 based on DNN-based beamforming with FastMNMF-guided adaptation.
FastMNMF (back end) is performed in a mini-batch style
 and the noisy and enhanced speech pairs are used together 
 with the original parallel training data 
 for updating the direction-aware DNN (front end) with backpropagation
 at a computationally-allowable interval.
This method is used with a blind dereverberation method called weighted prediction error (WPE) 
 for transcribing the noisy reverberant speech of a speaker,
 which can be detected from video or selected by a user's hand gesture or eye gaze,
 in a streaming manner and spatially showing the transcriptions with an AR technique. 
Our experiment showed that the word error rate was 
 improved by more than 10 points with the run-time adaptation using only twelve minutes observation.
\end{abstract}

\section{INTRODUCTION}

Sound source separation and enhancement form the basis of robot audition and computational auditory scene analysis~\cite{nakadai2017development,grondin2021odas}.
To comprehend and respond to the surrounding environment around robots or intelligent systems in real time, it is important to maintain both the low computational time and high separation performance.
Many real-time robot audition systems including HARK~\cite{nakadai2017development}, ODAS~\cite{grondin2021odas}, and ManyEars~\cite{grondin2013manyears} have been developed by combining multichannel source localization, separation, and recognition techniques.

Beamforming is an efficient multichannel source separation technique 
 that can extract the signal coming from a target direction 
 \cite{kumatani12micarray,vincent18book}.
There are various kinds of beamforming methods
 such as the delay-and-sum (DS) beamforming 
 and the minimum power distortionless response (MPDR) beamforming \cite{vincent18book},
 which require frequency-wise steering vectors 
 representing the sound propagation from the source to the microphones.
When pre-recorded steering vectors are used, 
 the separation performance is limited because
the actual environment may differ from the environment in which the steering vectors are recorded.
Other beamforming methods 
 such as the minimum variance distortionless response (MVDR) beamforming 
 and the generalized eigenvalue (GEV) beamforming
 \cite{souden10optimal,heymann17neuralbf,vincent18book}
 require the estimates of the spatial covariance matrices of the noise and target sources.

To estimate such spatial covariance matrices in beamforming,
 deep neural networks (DNNs) have been used 
 for estimating time-frequency (TF) masks of speech and noise
 from the observed mixture spectrogram, 
 sometimes given other features including directional information 
 \cite{heymann17neuralbf, Chen2018, Martin2019,Subramanian2020, Nakagome2020}.
Since DNNs are typically trained in a supervised manner,
 they are vulnerable to the mismatch of environmental conditions.
Conversely, multichannel blind source separation (BSS) methods
 such as multichannel non-negative matrix factorization (MNMF) \cite{ozerov09mnmf},
 independent low-rank matrix analysis (ILRMA) \cite{kitamura16ilrma}, and
 FastMNMF \cite{ito19fastmnmf,sekiguchi20fastmnmf},
 can work well in a variety of environments.
In particular,
 FastMNMF has been demonstrated to outperform MNMF and ILRMA \cite{sekiguchi20fastmnmf}.
Their computational cost, however,
 is relatively high, and thus they are not suitable for real-time processing.

\begin{figure*}[!ht]
\centering\includegraphics[width=.92\linewidth]{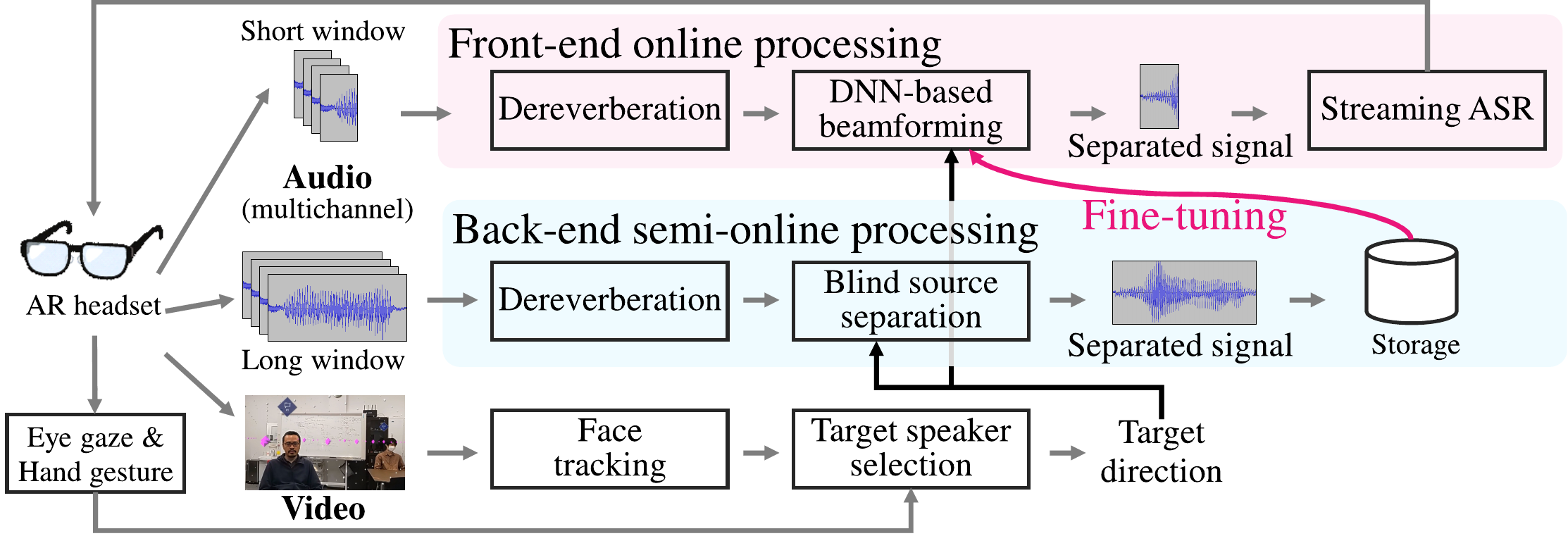}
\caption{The overview of the proposed system for online speech enhancement and recognition 
in real noisy echoic environments.}
\label{fig:overview}
\end{figure*}

To fuse the strengths of the supervised and unsupervised approaches
 in a practical real-time application,
 we propose a dual-process adaptive online speech enhancement framework consisting of
\textit{front-end} low-latency online DNN-based MVDR beamforming
 that provides enhanced speech for downstream tasks 
 (\eg, speech recognition and translation)
 and \textit{back-end} high-latency semi-online FastMNMF
 whose outputs are used for adapting the DNN to the actual environment.
To estimate the spectrogram of target speech,
 the demixing filters of MVDR beamforming are computed
 from speech and noise masks estimated by a DNN 
 that takes as input the observed mixture spectrogram and directional features.
The enhanced speech spectrogram given by FastMNMF, 
 together with the mixture spectrogram, 
 are used to update the DNN parameters in a supervised manner.

Those two components can be executed in parallel
 with separate computational resources.
Since the back-end FastMNMF and the DNN adaptation, which are based on iterative optimization,
 are computationally demanding,
 they are assumed to run with the support of graphic processing units (GPUs).
The front-end DNN-based beamforming can run 
 even on an edge device because of its relatively light-weight computation.

The proposed framework can be interpreted in terms of
 teacher-student learning \cite{hinton15distillation},
 where the front-end DNN acts as a student model 
 that tries to mimic the back-end FastMNMF acting as a teacher model.
Whereas BSS methods have conventionally been used
 for training a separation DNN from scratch
 in an offline manner
 \cite{Drude2019a,Bando2019a,nakagome20mentoring},
 our adaptation framework 
 uses the input and output (noisy and enhanced speeches) of FastMNMF 
 together with
 the original training data consisting of noisy and clean speech pairs
 for fine-tuning a pretrained separation DNN 
 at a computationally-allowable interval in a semi-online manner.

The downstream task we consider in this paper 
 is a streaming automatic speech recognition (ASR) \cite{yu21asr,lu21asr}
 for cocktail-party conversation assistance 
 with an augmented reality (AR) application with a head-mounted display (HMD).
This application can also be used
 for older adults and people with hearing impairment.
The proposed adaptive online speech enhancement framework 
 is expected to provide the accurate speech estimate with low latency
 for timely transcription,
 while being robust against acoustic conditions
 varying according to physical environments and source types and activities.
We thus integrate a blind dereverberation method called 
 weighted prediction error (WPE) \cite{yoshioka12wpe,drude18narawpe},
 which has been shown to be effective for improving the ASR performance 
 in diverse reverberant environments \cite{kinoshita16reverb},
 into both the front and back ends of the proposed framework.

In general, AR applications can help 
 a user perceive the surrounding environment
 by superimposing virtual contents on the user's view of the real world.
Adaptive systems are required 
 to provide the currently relevant contents
 based on the signals captured by various sensors attached to the HMD (AR headset).
AR applications have been studied in many different contexts, 
 \eg, health care \cite{geerse20gait,guinet21gait},
 medical care \cite{badiali20review,andrews21registration},
 manufacturing \cite{fragalamas18shipyard,avalle19fault},
 data analytics \cite{hoppenstedt19dataanalytics},
 and leisure \cite{picallo21arwireless}.
In addition to visual information and user poses
 that have been exploited in conventional studies,
 the ASR application considered also relies on audio information, 
 \ie, noisy reverberant audio signals.
The AR headset that we use is the Microsoft HoloLens 2 (HL2)\footnote{\url{https://www.microsoft.com/hololens/}},
 which is widely available in the market.
Using the data from the sensors attached on HL2,
 face tracking, eye tracking, and hand gesture recognition are performed
 to estimate the target speaker direction for guiding the speech enhancement.
To our knowledge, 
 speech enhancement for an AR headset has been dealt with only in \cite{donley21easycom},
 where beamforming is performed based on the target directions 
 given by a precise motion tracking system OptiTrack\footnote{\url{https://optitrack.com/}}
 whose markers are attached to AR headset prototype.
We experimentally evaluated the proposed system 
 using real recordings obtained by setting the HL2 on a dummy head.
For evaluation purpose only, we use an offline ASR system.
We confirmed that with the online adaptation of the DNN 
 using only twelve minutes of noisy observation,
 the average word error rate (WER)
 was improved by more than 10 points.

The rest of this paper is organized as follows.
Section II overviews our system for an AR headset with multi-modal sensors.
Section III describes the proposed direction-aware adaptive online speech enhancement framework.
Section IV presents the evaluation.
Section V concludes this paper.

\section{AUTOMATIC SPEECH RECOGNITION SYSTEM FOR AUGMENTED REALITY HEADSET}

This section provides an overview of our ASR system for AR headset
that incorporates the proposed adaptive online speech enhancement framework (Figure~\ref{fig:overview}).
At an abstract level, the system uses the visual information from the camera images to provide
possible target speakers from which the user can select using eye gaze or hand gesture.
The speech enhancement front-end, which is occasionally fine-tuned by the back-end, 
extracts possibly multiple speech signals
coming from the target directions corresponding to the user selection
and input the extracted speech signals into an ASR system.
The generated transcriptions are then displayed as virtual contents on the AR headset.

The rest of this section describes
the equipment used to realize the system,
the data communication,
the visual information processing,
the ASR system,
and the feedback to and from the AR headset user.
The proposed adaptive online speech enhancement framework is presented in the following section.

\subsection{Equipment}

The AR headset that we use is the Microsoft HoloLens 2 (HL2).
HL2 is an optical see-through HMD that allows the user to see the real-world surrounding with his/her own eyes,
not through cameras as in a video see-through HMD.
It has inertial measurement unit (IMU),
different types of cameras that are used for different purposes,
including head tracking and eye tracking,
and a 5-channel microphone array.
Among several AR headsets available in the market nowadays,
we choose HL2 not only because it has the sensors we need,
but also because it is supported by
the Mixed Reality Toolkit (MRTK)\footnote{https://docs.microsoft.com/windows/mixed-reality/}
that enables us to have rapid AR application development using Unity\footnote{https://unity.com/}.

As the server, we use a desktop or laptop computer with sufficient GPU computing power.
We describe the relevant specifications of the machine used for our experiments in Section \ref{sec:exp}.

\subsection{Data Communication}

Data is exchanged via Robot Operating System (ROS) \cite{quigley09ros}.
HL2 and the server are connected on the same wireless local area network so the HL2 user has freedom to move.
Wired connection can also be used for having a more reliable connection while charging the HL2,
\eg, in the case of recording large amount of data for the experiments.

HL2 sends
RGB (red, green, blue) images with a size of 960 x 540 pixels and 5 frames per second,
5-channel audio signals sampled at 48 kHz,
and information whether face indicators are selected by the user.
Different modules running on the server receive different parts of the data published by HL2.
The face tracking module takes the images to obtain the face positions such that
the program running on HL2 can generate selectable virtual indicators on top of the detected faces.
The speech enhancement module uses the selected face indicator information and the multichannel audio signals from HL2 to obtain the target speech signals for the ASR module inputs.
The ASR module sends the transcriptions to HL2 for which virtual contents are generated by the application running on HL2.

\subsection{Visual Information Processing}

\begin{figure}[!t]
\centering\includegraphics[width=.98\linewidth]{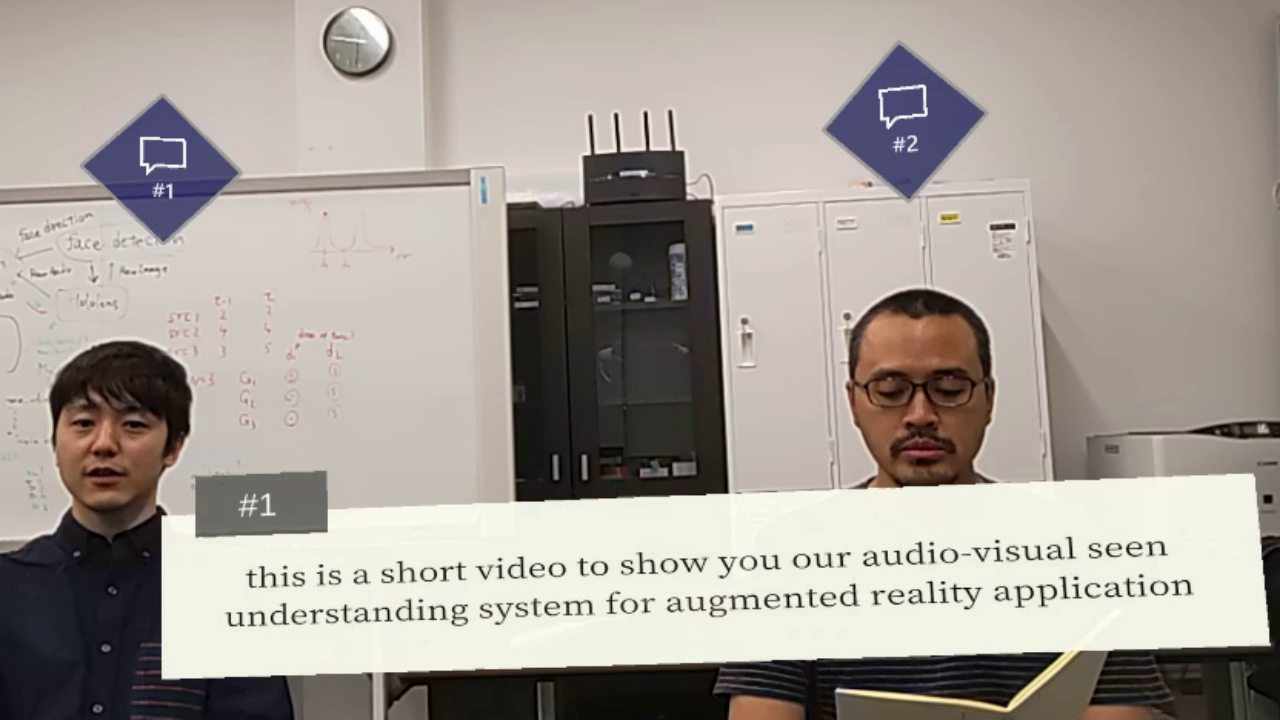}
\caption{An example of what the AR headset user sees.}
\label{fig:user_pov}
\end{figure}

The AR headset user's eyes and hands are tracked on the HL2 using MRTK.
The face tracking based on FaceNet \cite{schroff15facenet} is done on the server given the RGB images from HL2.
The resulting bounding boxes on 2-dimensional images are then projected to the 3-dimensional virtual world
using the camera transformation matrices that are also sent by HL2
and the ratio of the detected face width to a pre-defined face width reference
such that virtual face indicators can be generated on top of the detected face (see Figure~\ref{fig:user_pov}).

\subsection{Automatic Speech Recognition}

The speech transcription is supposed to be obtained
using a streaming ASR system \cite{yu21asr,lu21asr}, such as one based on
Microsoft Azure\footnote{\url{https://azure.microsoft.com/services/cognitive-services/speech-to-text/}}
or Google Cloud\footnote{\url{https://cloud.google.com/speech-to-text/}} service.
Using a streaming ASR, the user can continuously receive transcriptions.
The latency of the ASR system is beyond the scope of this paper.

\subsection{Human-Machine Interaction}

Figure~\ref{fig:user_pov} illustrates what the AR headset user sees.
The face indicators and the text panel showing the ASR transcriptions are contents generated in the virtual world.
The face indicators are virtually placed on top of the real-world detected face.
The indicators are also served as pressable buttons for selecting target speakers.
The user can focus on a face indicator using his/her eye gaze, and then select the indicator using a pinching hand gesture.
The text panel is always in front of the user following his/her movement to make sure that the text is legible.

\section{DIRECTION-AWARE ADAPTIVE ONLINE SPEECH ENHANCEMENT FRAMEWORK}\label{sec:proposed}

This section explains in detail the proposed adaptive online speech enhancement framework
that composed of a front-end online processing based on beamforming with a separation DNN
and a back-end semi-online processing based on multichannel BSS for fine-tuning the front-end DNN.
Both front-end and back-end systems use WPE for dereverberation \cite{yoshioka12wpe,drude18narawpe}.

Let $\mX \triangleq \{\vx_{ft}\}_{f,t=1}^{F, T} \in \mathbb{C}^{F \times T \times M}$ be the observed multichannel mixture short-time Fourier transform (STFT) spectrogram, 
 where $F$, $T$, and $M$ represent the numbers of frequency bins, time frames, and microphones, respectively.
Both front-end and back-end parts apply the block-online processing,
where one block is composed of a sequence of multiple time frames.
To have outputs with a low latency,
the front-end part can use
a smaller block size to reduce the computational time and 
a smaller block shift size to perform computations more often.
Conversely, 
the back-end part can use
a larger block size to obtain more data useful for BSS.

\subsection{Front-end DNN-Based Beamforming} \label{sec:frontend}

\begin{figure*}[!t]
\centering\includegraphics[width=.75\linewidth]{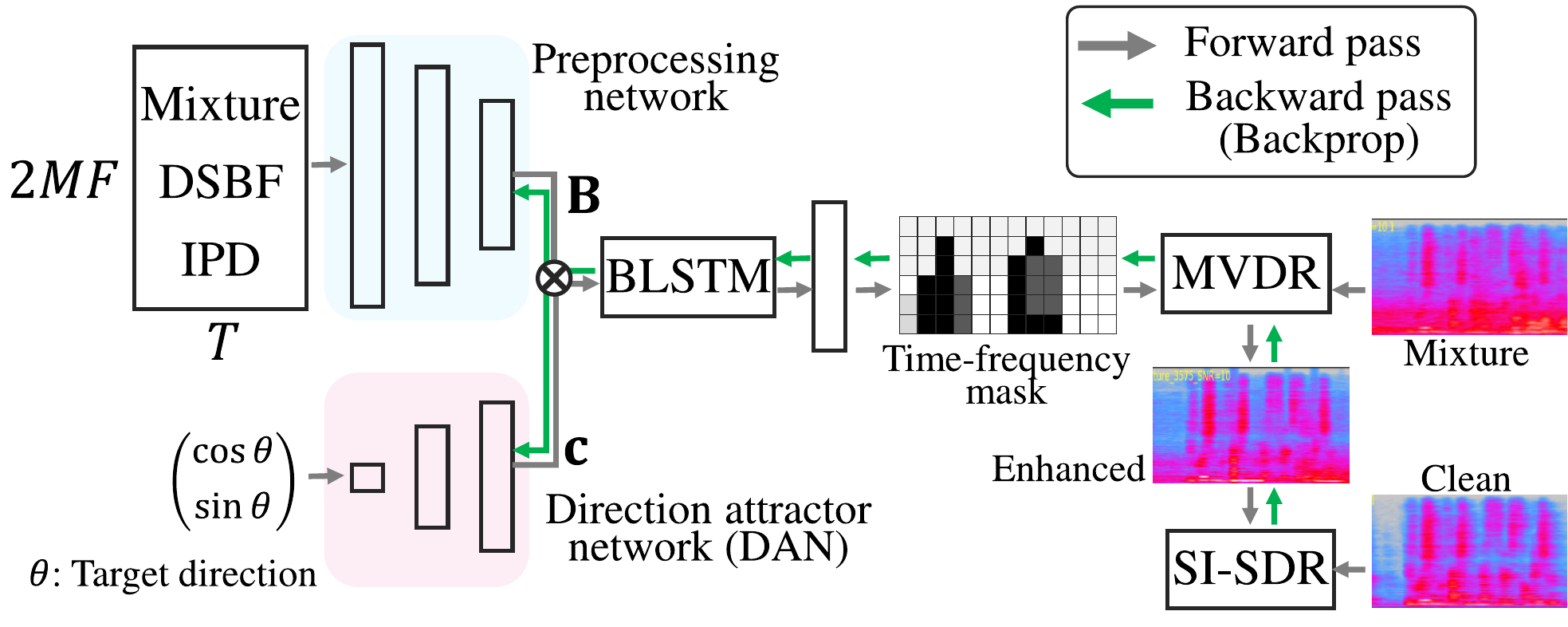}
\vspace{-3mm}
\caption{The architecture of the front-end DNN-based beamforming.}
\label{fig:frontend}
\vspace{1mm}
\end{figure*}

Given multichannel mixture spectrogram and spatial information about a target direction as input,
the separation network estimates the time-frequency (TF) masks of the signal coming from the target direction (Figure~\ref{fig:frontend}).
Specifically, the inputs are
the log-magnitude spectrograms of first channel of the mixture signals,
the log-magnitude spectrograms of the delay-and-sum beamforming (DSBF) \cite{vincent18book} results,
inter-channel phase differences (IPDs) expressed as the cosine and sine of $\{ \angle \frac{x_{ft,m(\ge2)}}{x_{ft1}} \}_{f=1,t=1,m=2}^{F,T,M}$,
and an angle of the target direction.
For each block composed of $T$ frames,
a matrix of size $T \times (2MF)$ consisting of the log-magnitude spectrograms and IPDs is fed into a preprocessing network that outputs a matrix $\mB = [\vb_1, \ldots, \vb_T]^\Tr \in \mathbb{R}^{T \times L}$, 
where $L$ is the input dimension of the following network.
The cosine and sine of the angle of the target direction are fed into a direction attractor network (DAN) \cite{Nakagome2020} to obtain a vector $\vc \in \mathbb{R}^L$.
$\{ \vb_t \}_{t=1}^T$ and $\vc$ are then multiplied in an element-wise manner
and fed into a bidirectional long short-term memory (BLSTM) network with
a final fully-connected layer resulting in TF masks.

The estimated TF masks $\{m_{ft}\}_{f,t=1}^{F,T}$ are used for calculating a minimum variance distortionless response (MVDR) beamformer \cite{souden10optimal} $\vw_{f}^{\text{MVDR}}$ given as follows:
\begin{gather}
    \vw_{f}^{\text{MVDR}} = \frac{
            \bm{\Sigma}_{\mathbf{N},f}^{-1} \bm{\Sigma}_{\mathbf{S},f}^{\vphantom{-1}}
        }{ \text{tr} \left(
                \bm{\Sigma}_{\mathbf{N},f}^{-1} \bm{\Sigma}_{\mathbf{S},f}^{\vphantom{-1}}
            \right)
        }   \mathbf{u}_{n_0} ,
\end{gather}
where $\mathbf{u}_{n_0}$ is a one-hot vector whose ${n_0}$-th entry is $1$ and $0$ elsewhere and
${n_0}$ is the index of the reference microphone (\eg, the top center microphone of HL2).
$\bm{\Sigma}_{\mS,f}$ and $\bm{\Sigma}_{\mN,f}$ are speech and noise spatial covariance matrices (SCMs), respectively, and calculated as follows:
\begin{align}
    \bm{\Sigma}_{\mS,f} &= \sum_{t=1}^T m_{ft} \vx_{ft} \vx_{ft}^\Hr , \\
    \bm{\Sigma}_{\mN,f} &= \sum_{t=1}^T (1 - m_{ft}) \vx_{ft} \vx_{ft}^\Hr .
\end{align}
A source SCM encodes the spatial characteristics (including direction and spread) of a source as seen by the microphone array \cite{duong10underdetermined}.
The enhanced time-frequency domain signal given by $(\vw_{f}^{\text{MVDR}})^\Hr \vx_{ft}$ is transformed 
 into a time-domain signal $\vs_\text{enh}$ by using inverse STFT.
In the pre-training step, negative SI-SDR \cite{leroux19sisdr} between the enhanced signal $\vs_\text{enh}$ and the clean signal of the target direction $\vs_\text{ref}$ is used as a loss function:
\begin{align}
    \text{SI-SDR}(\vs_\text{enh}, \vs_\text{ref}) &= 10 \log_{10}
    \left(
        \frac{ ||\alpha \vs_\text{ref}||^2}
        { || \alpha \vs_\text{ref} - \vs_\text{enh}||^2}
    \right),
\end{align}
where $\alpha = \vs_\text{enh}^\Tr \vs_\text{ref} / ||\vs_\text{ref}||^2$ is a scaling factor.
In the fine-tuning step, instead of the clean reference signals, the outputs of the back-end BSS are used as reference.

\begin{figure*}[!t]
\centering\includegraphics[width=.95\linewidth]{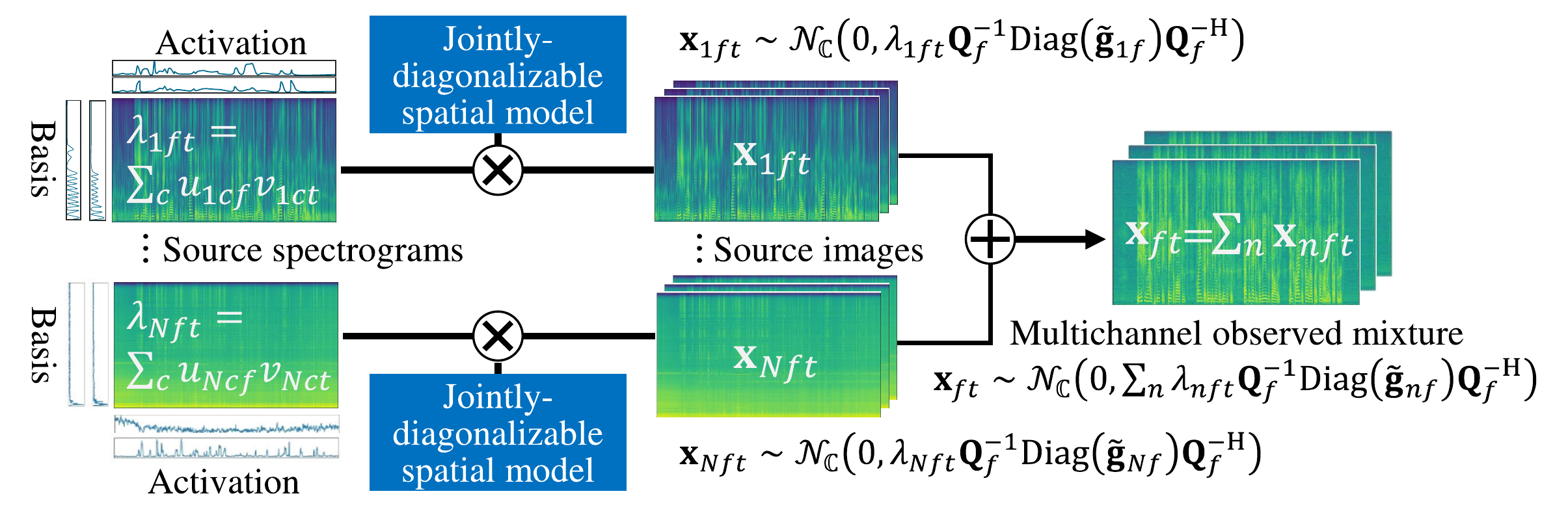}
\vspace{-3mm}
\caption{The probabilistic generative model of a multichannel mixture using a jointly-diagonalizable full-rank spatial model \cite{sekiguchi20fastmnmf}.}
\label{fig:fastmnmf}
\end{figure*}

\subsection{Back-end Blind Source Separation} \label{sec:backend}

A widely-used approach to BSS is 
 to formulate and optimize a probabilistic model 
 of the observed multichannel mixture spectrogram $\mX$ 
 consisting of a spatial model that represents the SCMs
 and a source model that represents the power spectral densities (PSDs).
Our system adopts the state-of-the-art BSS method called FastMNMF \cite{sekiguchi20fastmnmf} shown in Figure \ref{fig:fastmnmf}
 that consists of a jointly-diagonalizable (JD) spatial model and 
 a nonnegative matrix factorization (NMF)-based source model.

Suppose that a mixture of $N$ sources is recorded by $M$ microphones.
Let $\mX_n \triangleq \{\vx_{nft}\}_{f,t=1}^{F,T}$ be the multichannel spectrogram of source $n$ called image.
FastMNMF assumes that each source image $\mathbf{x}_{nft}$ follows
    an $M$-variate complex-valued circularly-symmetric Gaussian distribution,
    whose covariance matrix is jointly-diagonalizable by a time-invariant \textit{diagonalization matrix} shared among all sources $\mathbf{Q}_f \!\in\! \smash{\mathbb{C}^{M \!\times\! M}}$:
\begin{equation}
\mathbf{x}_{nft} \sim
\mathcal{N}_\mathbb{C}^{M} \left(
    \mathbf 0,
    \lambda_{nft} \mathbf{Q}_f^{-1} \mathrm{Diag}(\tilde{\mathbf{g}}_{n}) \mathbf{Q}_f^{-\mathsf{H}}
\right) ,
\end{equation}
    where $\lambda_{nft} \!\in\! \mathbb{R}_{+}$ is the PSD,
    and $\mathrm{Diag}(\tilde{\mathbf{g}}_{n})$ is a diagonal matrix
    whose diagonal elements $\tilde{\mathbf{g}}_{n} \!\triangleq\! \smash{\left[ \tilde{g}_{1n}, \dotsc, \tilde{g}_{Mn} \right]^\mathsf{T}} \!\in\! \mathbb{R}_{+}^{M}$
    is a frequency-invariant nonnegative vector \cite{sekiguchi20fastmnmf}. 
This joint-diagonalization leads to the mixture decorrelation:
\begin{equation}
\mathbf{y}_{ft} \triangleq \mathbf{Q}_f \mathbf{x}_{ft} \sim
\mathcal{N}_\mathbb{C}^{M} \left(
    \mathbf 0, \sum_{n=1}^{N} \lambda_{nft} \mathrm{Diag}(\tilde{\mathbf{g}}_{n})
\right) \label{eq:decorr_fastmnmf},
\end{equation}
so $\{ \smash{\mathbf{y}_{ft} \}_m} \!\sim\! \smash{\mathcal{N}_\mathbb{C} \big(
    0, \smash{\sigma^2_{mft}} \!\triangleq\!
    \smash{\sum_{n=1}^{N} \lambda_{nft} \tilde{g}_{mn} } \big)}$
are independent of each other.
We optimize the parameters
$\mathbf{\Phi} \!\triangleq\! \{ \mathbf{Q}_f, \allowbreak u_{ncf}, \allowbreak v_{nct}, \allowbreak \tilde{g}_{mn} |\allowbreak \forall m, \forall n, \forall f, \forall t, \forall c \}$
by maximizing
\begin{gather}
\ln p(\mathbf{X}) =
    \sum_{f,t,m=1}^{F,T,M} \ln p(\{\mathbf{y}_{ft} \}_m)
    + T \sum_{f=1}^{F} \ln \left| \mathbf{Q}_f \right|^{2} ,
    \label{eq:fastmnmf_ll} \\
\ln p( \{\mathbf{y}_{ft} \}_m) =
    - 
        \frac{\left| \{ \mathbf{y}_{ft} \}_m \right|^2}
        {\sigma^2_{mft}}
        - \ln \sigma^2_{mft}
    + \text{const.} \label{eq:fastmnmf_ll_y}
\end{gather}
The estimated source image $\widehat{\mathbf{x}}_{nft}$
can then be computed given $\mathbf{x}_{ft}$ and $\mathbf{\Phi}$
by Wiener filtering:
\begin{equation}
\widehat{\mathbf{x}}_{nft} =
    \mathbf{Q}_f^{-1}
    \mathrm{Diag} \left(
        \frac{\lambda_{nft} \tilde{\mathbf{g}}_{n}}
        {\sum_{n^{\prime}=1}^{N} \lambda_{n^{\prime}ft}\tilde{\mathbf{g}}_{n^{\prime}}} \right)
    \mathbf{Q}_f \mathbf{x}_{ft} .
    \label{eq:wiener_filtering}
\end{equation}

The parameter optimization is performed in an iterative manner, and
sophisticated initialization often improves the performance and makes convergence speed faster.
In the beginning of the iterations, a frequency-invariant source model given by $\lambda_{nft} = \lambda_{nt}$ is used
to alleviate the frequency permutation problem. 
Then, the source model is switched to an NMF-based model given by $\lambda_{nft} = \sum_{c=1}^C u_{ncf} v_{nct}$ to precisely represent the PSDs,
where $u_{ncf} \!\in\! \mathbb{R}_{+}$ is a basis,
$v_{nct} \!\in\! \mathbb{R}_{+}$ is an activation,
and $C$ is the number of NMF components.

In our system, the direction of the target sound source obtained from the visual information is used for initialization.
Because the column vectors of $\mQ_f^{-1}$ in the JD spatial model are known to play a similar role as steering vectors \cite{sekiguchi20fastmnmf},
 the first column vector of $\mQ_f^{-1}$ is initialized to the steering vector $\va_f$ of the target direction,
 and $\tilde{\vg}_1$ is initialized to a one-hot-like vector whose first element is 1 and other elements are $\epsilon$ (\eg, $0.01$).
This means that the SCM of source $1$ is set to a matrix that is close to the rank-1 matrix $\va_f \va_f^\Hr$.
Note that the target sound source is not always active,
 and if it is not active, the SCM will move away from $\va_f \va_f^\Hr$.

For storing fine-tuning data,
 we need to estimate which estimated source image corresponds to the target direction.
Thanks to the initialization explained above, source $1$ often corresponds to the target direction.
However, this does not always hold, 
 especially when the steering vector used for initialization significantly differs from the actual steering vector 
 and when the target source is inactive.
This calls for the estimation of the direction for each source image.

Inspired by the source localization method called multiple signal classification (MUSIC) \cite{schmidt86music},
we estimate the correspondences between source images and directions
using the eigenvectors $\{ \vv_{nfm} \}_{m=1}^M$ computed from the source SCM $\mG_{nf} \triangleq \mathbf{Q}_f^{-1} \mathrm{Diag}(\tilde{\mathbf{g}}_{n}) \mathbf{Q}_f^{-\mathsf{H}}$.
Assuming that source $n$ corresponds to the target direction, 
 the principal eigenvector $\vv_{nf1}$ is close to the steering vector of the target direction $\va_f$,
 and the other eigenvectors $\{ \vv_{nfm} \}_{m=2}^M$ are orthogonal to the steering vector.
Thus, the degree of response given by 
 $l_n \triangleq \sum_{f=1}^F \sum_{m=2}^M | \va_f^\Hr \vv_{nfm} |^2 $ becomes small for the target direction
 and large for other directions.
If $l_n$ is smaller than a threshold, 
 source $n$ is regarded as the target source,
 and the corresponding estimated source image 
 is stored in a buffer for fine-tuning the separation DNN.
 
\subsection{Online Fine-tuning Strategy} \label{sec:online_finetune}
To collect data for fine-tuning the front-end separation network, 
 the BSS method is always working in the back-end.
 If the estimated source image corresponding to the target direction is obtained,
 the source image, the target direction, and the observed multichannel signals are stored.
When the amount of the data exceeds a threshold, fine-tuning of the separation network is executed.
To reduce the computational cost and to deal with the change of a surrounding environment, only recent data is used for fine-tuning.
The first fine-tuning step updates the pre-trained network parameters, while the subsequent fine-tuning steps update the network parameters obtained from the previous fine-tuning step.

\section{EVALUATION AND DISCUSSION}\label{sec:exp}
In this section, we evaluate the effectiveness of the online fine-tuning of the separation DNN by using data recorded with Microsoft HoloLens 2 (HL2).

\begingroup 
\tabcolsep=3pt
\renewcommand{\arraystretch}{1.2}
  \begin{table}[!t]
  \caption{Average WERs [\%] and computation times [s] of the baseline methods.
  The total time latency [s] is the sum of the block shift size and the average computation time for each block.}
  \vspace{-\baselineskip}
  \label{table:exp_baseline}
  \begin{center}
    \scalebox{1.0}{
    \begin{tabular}{|c|c|c|c|c|c|}
        \hline
        & \multicolumn{2}{c|}{Block} & Comp. & Total & WER \\
        \cline{2-3}
        Method & Size [s] & Shift [s] & Time [s] & Latency [s] & [\%] \\
        \hline
        Clean & -- & -- & -- & -- & 6.1 \\
        Observation (Noisy) & -- & -- & -- & -- & 92.1 \\
        \hline
        DSBF & 3.07 & 0.51 & 0.04 & 0.55 & 70.5 \\
        MPDR & 3.07 & 0.51 & 0.04 & 0.55 & 53.0 \\
        FastMNMF & 3.07 & 0.51 & 0.66 & 1.17 & 29.2 \\
        FastMNMF & 3.07 & 3.07 & 0.68 & 3.75 & 24.2 \\
        FastMNMF & 9.22 & 9.22 & 1.89 & 11.11 & 15.0 \\
        \begin{tabular}{c} Pre-trained DNN \\ + MVDR \end{tabular} & 3.07 & 0.50 & 0.25 & 0.75 & 35.6 \\
        \hline
      \end{tabular}
    }
  \end{center}
  \vspace{-1mm}
\end{table}
\endgroup

\subsection{Experimental Conditions}
For fine-tuning and evaluation, we separately recorded 5-channel ($M=5$) reverberant speech signals and diffuse noise in a room with an $\text{RT}_{60}$ of about 800 ms by using a HL2.
The HL2 was set on a dummy head,
 and a loudspeaker was placed at each of eight directions with 45 degrees interval.
The distances between the HL2 and the loudspeakers were set to 1.5 meters except for 135 degrees,
 and that at 135 degrees was 3 meters.
The loudspeaker at 0 degree was regarded as the target, and that of the other directions except for 135 degrees were regarded as interference speakers.
The loudspeaker at 135 degrees were used for noise.
To make the noise signals diffuse, a room partition was placed between a loudspeaker and the device, 
 and the loudspeaker was directed to the opposite direction to the device.
The dry speech signals were randomly selected from the {\it test-clean} subset of Librispeech dataset \cite{panayotov15librispeech},
and the noise signals were randomly selected from the {\it backgrounds} folder in CHiME3 dataset \cite{vincent17chime4}.
We then synthesized 66 minutes noisy mixture signals consisting of two reverberant speech and one diffuse noise signals.
The signals were split into 48 minutes \textit{fine-tuning} data and 18 minutes \textit{evaluation} data.

For pre-training the separation network,
 we synthesized noisy mixture signals of about 15 hours consisting of two speech and diffuse noise signals by using recorded impulse responses.
The dry speech signals were randomly selected from WSJ0 SI-84 training set, 
and the noise signals were randomly selected from DEMAND dataset.
The impulse responses were recorded at 72 directions with 5 degrees interval in an anechoic room.
To simulate diffuse noise signals, 
 we summed up 36 different noise signals each of which was convolved with one of 36 impulse responses with 10 degrees interval.

Audio signals were sampled at 16 kHz and processed by STFT with 
 a Hann window of 1024 points ($F=513$) and a shifting interval of 256 points.
The block size of the back-end FastMNMF for making the fine-tuning data was 561 frames (about 9 seconds), and the number of NMF components $C$ was 8.
The number of iterations for updating FastMNMF with the frequency-invariant source model was 50, 
 and that for updating vanilla FastMNMF was 50.
For both front-end and back-end, WPE with the tap length of 5, the delay of 3, and the number of iterations of 3 was used.

The block size of the separation network was 189 frames (about 3 seconds).
As preprocessing network and direction attractor network described in Section~\ref{sec:frontend}, we used three fully-connected layers,
 and the dimension $L$ was 1024.
The number of layers of the BLSTM was three and the dimension of the hidden layers was 512.
In the pre-training step, the separation network was trained with a learning rate of 0.001, batch size of 128, and the maximum number of epochs of 500.
In the inference step, it was used with a shifting interval of 8000 points (0.5 seconds), 
 and except for the first iteration, only the last 8000 points were used as output.
In the fine-tuning step, the network was trained with a learning rate of 0.001 and batch size of 16.
To stabilize the training, in addition to the fine-tuning data, we also used the data used in the pre-training step.
The ratio between the fine-tuning and pre-training data was set to 1:1.

To confirm the upper bound of the fine-tuning,
 we tested off-line fine-tuning that used a part of the fine-tuning data at one time.
The amount of data was $\{3, 6, 12, 18, 24, 30, 36, 42, 48\}$ minutes, and the maximum number of epochs was 50.
Then, we tested online fine-tuning discussed in Section~\ref{sec:online_finetune}.
The fine-tuning was executed every 3 minutes, and the amount of data used in one fine-tuning was 12 minutes at most.
Until 12 minutes of data was obtained, all of the accumulated data was used.
After that, only the latest 12 minutes of data was used.
As for the number of epochs per one fine-tuning, we tested $\{1, 3, 5, 10\}$.

\begingroup
\tabcolsep=3.5pt
\renewcommand{\arraystretch}{1.2}
  \begin{table}[!t]
  \caption{Average WERs [\%] of the offline fine-tuning}
  \vspace{-\baselineskip}
  \label{table:exp_offline}
  \begin{center}
    \scalebox{1.0}{
    \begin{tabular}{|c|c|c|c|c|c|c|c|c|c|}
        \hline
         \multirow{2}{*}{\begin{tabular}{c} Fine-tune \\ Data [min] \end{tabular}} & \multicolumn{9}{c|}{The number of epochs for fine-tuning} \\
        \cline{2-10}
         & 0 & 1 & 3 & 5 & 10 & 20 & 30 & 40 & 50 \\
        \hline
         3 & 35.6 & 31.6 & 30.8 & \bf{29.2} & 31.6 & 34.0 & 32.8 & 34.9 & 35.3 \\
         6 & 35.6 & 27.3 & 29.4 & 28.0 & 28.4 & \bf{26.8} & 30.8 & 31.7 & 31.3 \\
        12 & 35.6 & 26.4 & 25.3 & \bf{23.6} & 27.0 & 26.9 & 29.2 & 27.9 & 28.1 \\
        18 & 35.6 & 25.7 & 26.9 & \bf{24.1} & 25.6 & 26.9 & 24.2 & 26.5 & 26.4 \\
        24 & 35.6 & 26.8 & \bf{22.9} & 26.8 & 28.2 & 25.8 & 24.9 & 24.2 & 24.9 \\
        30 & 35.6 & 25.6 & 24.0 & \bf{21.3} & 23.7 & 23.0 & 22.5 & 22.9 & 21.9 \\
        36 & 35.6 & 25.1 & 23.0 & 23.7 & \bf{21.0} & 22.0 & 23.9 & 22.3 & 23.4 \\
        42 & 35.6 & 26.1 & 27.7 & 20.8 & 22.5 & \bf{20.4} & 22.2 & 25.5 & 25.6 \\
        48 & 35.6 & 23.7 & \bf{21.3} & 25.2 & 23.3 & 21.7 & 22.9 & 24.1 & 28.6 \\
         \hline
      \end{tabular}
    }
  \end{center}
\end{table}
\endgroup

As a baseline, we tested delay-and-sum beamforming (DSBF), minimum power distortionless response (MPDR) beamforming, and FastMNMF with block size of 192 frames (about 3 seconds) and shift size of 32 frames (about 0.5 seconds), which was almost the same as the front-end DNN-based beamforming.

The word error rate (WER) was used as a criterion for evaluating the performance.
For evaluation purpose only, we used an offline ASR system using the transformer-based acoustic and language models of the SpeechBrain toolkit \cite{ravanelli21speechbrain}.
The computation time for processing one block was measured on Intel Xeon W-2145 (3.70 GHz) with NVIDIA GeForce Titan RTX.

\begin{figure}[!t]
\centering\includegraphics[width=.99\linewidth]{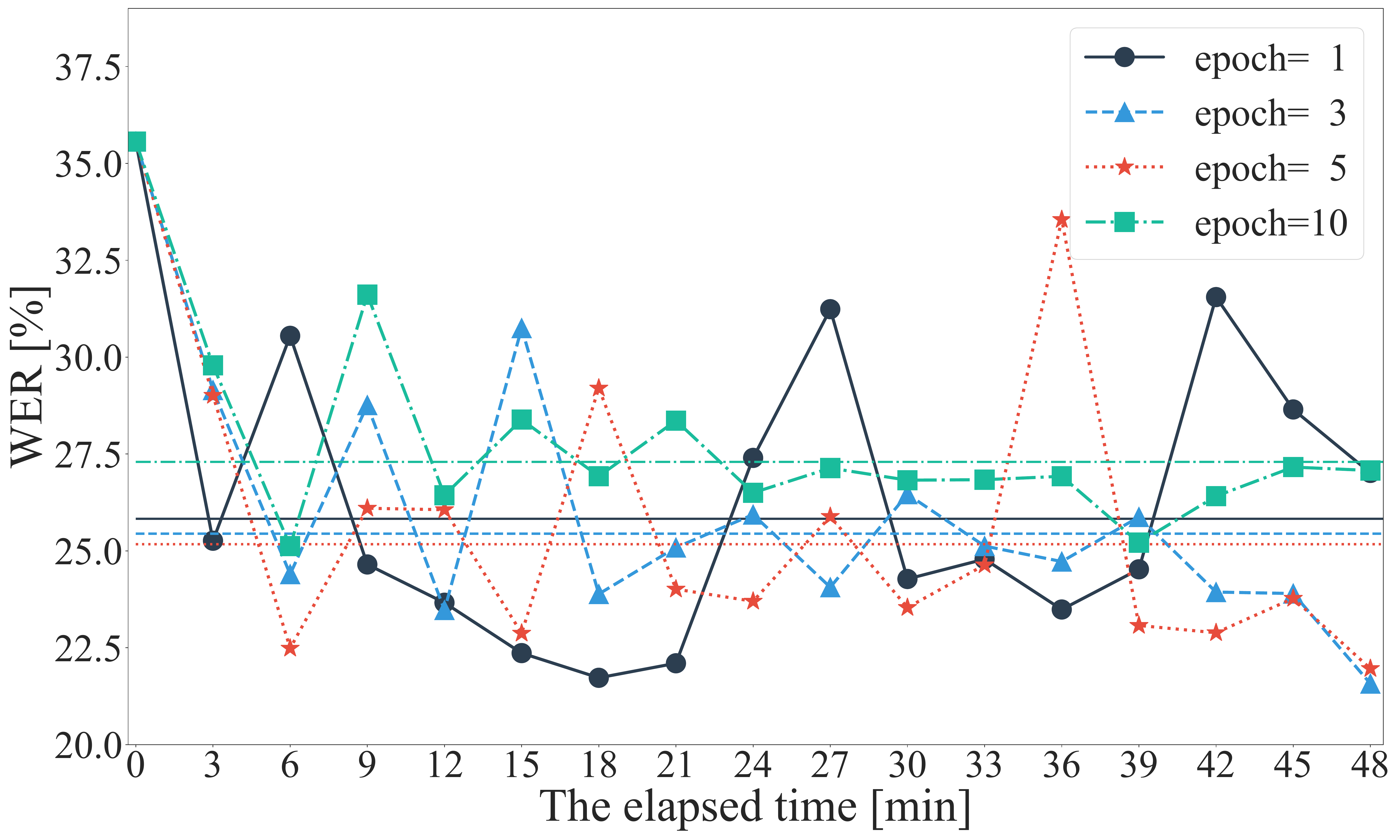}
\caption{Average WERs [\%] of the online fine-tuning. The horizontal lines indicate the average WERs over all time.}
\label{fig:exp_online}
\end{figure}

\subsection{Experimental Results}

Table~\ref{table:exp_baseline} shows the average WERs and computation times of the baseline methods, and Table~\ref{table:exp_offline} shows the average WERs of the front-end DNN-based beamforming with the offline fine-tuning, where the base performance using the pre-trained DNN is shown when the number of epochs equals zero.

In terms of the ASR performance, FastMNMF outperformed the other baseline methods and the DNN-based beamforming using a pre-trained DNN.
With a larger block size, the performance of FastMNMF was improved, but the computation time was also increased.
Even with a block size of about 3 s and a block shift size of about 0.5 s, the computation time of 0.66 s was more than the block shift, making real-time processing impossible to achieve the same ASR performance level.
Conversely, using the front-end DNN-based beamforming, the average computation time for processing one block with a similar size was only 0.25 s.
These experimental results validate that FastMNMF is not suitable for front-end processing.

Table~\ref{table:exp_offline} further demonstrates that FastMNMF is useful as a back-end for fine-tuning the front-end DNN.
After fine-tuning for 10 epochs or less, the best performance can be achieved in almost all conditions.
When the amount of data is small, too many epochs caused overfitting and decreased the performance.
With only 12 minutes of data, the WER improved 12 points from that without fine-tuning,
 and it is better than that of FastMNMF with almost the same block size (3 seconds).
With 42 minutes of data, the WER improvement was the highest (15.2 points).

\begingroup
\tabcolsep=2.5pt
\renewcommand{\arraystretch}{1.2}
  \begin{table}[!t]
  \caption{Computation time [s] of fine-tuning per one epoch}
  \vspace{-\baselineskip}
  \label{table:exp_comp_time}
  \begin{center}
    \scalebox{1.0}{
    \begin{tabular}{|c|c|c|c|c|c|c|c|c|c|}
        \hline
        Fine-tune Data [min] & 3 & 6 & 12 & 18 & 24 & 30 & 36 & 42 & 48 \\
        \hline
        Comp. Time [s] & 11.4 & 21.7 & 42.5 & 63.3 & 84.4 & 105 & 126 & 146 & 169 \\
        \hline
      \end{tabular}
    }
  \end{center}
\end{table}
\endgroup

Figure~\ref{fig:exp_online} shows the average WERs of the online fine-tuning.
The average WERs over all time were 25.8, 25.4, 25.2, and 27.3\% when the number of epochs was 1, 3, 5, and 10, respectively.
Although the performance sometimes fluctuated,
 the average WERs improved by more than 10 points from that without the fine-tuning.
Table~\ref{table:exp_comp_time} shows that performing fine-tuning with 12 minutes of data for 1 or 3 epochs is a reasonable choice because the total computation time is less than the fine-tuning interval duration, i.e., 3 minutes.
The computation time can be further reduced by changing the ratio between the fine-tuning and original training data,
 although it sometimes makes the training unstable.

\section{CONCLUSION}
In this article, we proposed the streaming ASR system 
 with the front-end DNN-based beamforming and the back-end semi-online BSS for fine-tuning the front-end DNN.
Using visual information and eye gaze and hand gesture recognition, the device can select a target speaker,
 and the transcription of the target speaker given from the front-end system is continuously shown on the device.
To compensate for the sensitivity to a surrounding environment of the front-end DNN,
 the back-end system accumulates the signals from the target speaker by using the state-of-the-art BSS method,
 and periodically fine-tunes the DNN.
In the experiment, we confirmed that the WER improves by 12 points with only twelve minutes of fine-tuning data.

\IEEEtriggeratref{29}
\bibliographystyle{./IEEEtran}
\bibliography{./IEEEabrv,./MYabrv,./refs}

\end{document}

%% file: definition.tex
\usepackage{amsmath,amsfonts,amssymb,bm}

\newcommand{\ie}{i.e.}
\newcommand{\eg}{e.g.}

\newcommand{\Tr}{\mathsf{T}}
\newcommand{\Hr}{\mathsf{H}}

\newcommand{\mB}{\mathbf{B}}

\newcommand{\mG}{\mathbf{G}}

\newcommand{\mN}{\mathbf{N}}

\newcommand{\mQ}{\mathbf{Q}}

\newcommand{\mS}{\mathbf{S}}

\newcommand{\mX}{\mathbf{X}}

\newcommand{\va}{\mathbf{a}}
\newcommand{\vb}{\mathbf{b}}
\newcommand{\vc}{\mathbf{c}}

\newcommand{\vg}{\mathbf{g}}

\newcommand{\vs}{\mathbf{s}}

\newcommand{\vv}{\mathbf{v}}
\newcommand{\vw}{\mathbf{w}}
\newcommand{\vx}{\mathbf{x}}